\documentclass[showpacs,aps,prc,longbibliography,nofootinbib,showkeys,superscriptaddress,twocolumn,reprint]{revtex4-1}

\usepackage[colorlinks, urlcolor=blue,linkcolor=blue, anchorcolor=blue, citecolor=blue]{hyperref}
\usepackage[english]{babel}
\usepackage[utf8]{inputenc}
\usepackage{multirow}
\usepackage{array}
\usepackage{amsthm}
\usepackage{mathtools}
\usepackage{physics}
\usepackage{xcolor}
\usepackage{graphicx}
\usepackage{bm}
\usepackage{soul}
\usepackage{adjustbox}
\usepackage{placeins}
\usepackage[T1]{fontenc}
\usepackage{lipsum}
\usepackage{csquotes}
\usepackage{float}
\usepackage{booktabs}
\usepackage{microtype}

\graphicspath{{figs/}}

\newcommand{\trento}{{\sc TRENTo}}
\newcommand{\music}{{\sc MUSIC}}
\newcommand{\iss}{{\sc iSS}}
\newcommand{\SMASH}{{\sc SMASH}}
\newcommand{\snn}{\sqrt{s_\mathrm{NN}}}

\newcommand{\ld}[1]{{\color{black} #1}}

\begin{document}

\title{Characterizing radial flow fluctuations in relativistic heavy-ion collisions\\ at top RHIC and LHC energies}

\author{Lipei Du}
\affiliation{Department of Physics, University of California, Berkeley CA 94270}
\affiliation{Nuclear Science Division, Lawrence Berkeley National Laboratory, Berkeley CA 94270}

\date{\today}
\begin{abstract}
This study presents a systematic investigation of the transverse-momentum differential radial flow fluctuations observable $v_0(p_T)$ in relativistic heavy-ion collisions at top RHIC ($\snn{\,=\,}200$~GeV) and LHC ($\snn{\,=\,}2.76$ and 5.02 TeV) energies. Using a multistage hydrodynamic model, this study assesses the sensitivity of $v_0(p_T)$ to a wide range of physical effects, including bulk and shear viscosities, off-equilibrium corrections at particlization, the presence of a hadronic afterburner, and the nucleon size in the initial conditions. By employing complementary rescaling strategies, this study demonstrates how different physical effects leave distinct imprints on the shape of $v_0(p_T)$. A combined double-rescaling of $v_0(p_T)/v_0$ versus $p_T/\langle p_T \rangle$ reveals a universality across a wide range of energies and model assumptions in the low-$p_T$ regime, a robust signature of collective behavior. This allows us to disentangle the universal dynamics of the bulk medium from model-specific features that emerge at higher $p_T$. These results establish $v_0(p_T)$ as a powerful and complementary observable for constraining QGP transport properties and initial-state granularity, offering a unique probe of the created QCD medium.
\end{abstract}

\maketitle

\section{Introduction}\label{sec:intro}

The collective expansion of the hot and dense medium created in relativistic heavy-ion collisions manifests in both anisotropic and isotropic components. The anisotropic flow~\cite{Ollitrault:1992bk, Voloshin:2008dg}, characterized by Fourier coefficients $v_n$ of the azimuthal particle distribution, has been instrumental in demonstrating the near-perfect fluidity of the quark-gluon plasma (QGP) and in constraining its transport properties, such as the shear viscosity to entropy density ratio $\eta/s$ \cite{Song:2010mg, Heinz:2013th, Bernhard:2019bmu}.

In addition to anisotropic flow, the system undergoes an isotropic expansion, known as \textit{radial flow}, which pushes particles to higher transverse momenta $p_T$. Traditionally, radial flow has been studied by analyzing the shapes of $p_T$ spectra, often using blast-wave models to extract an average radial velocity $\langle \beta_T \rangle$~\cite{Schnedermann:1993ws,Tang:2008ud}.\footnote{Recent studies have proposed using electromagnetic tomography to probe early-time radial flow in heavy-ion collisions \cite{Du:2024pbd,Du:2025dot}. This method combines expansion-insensitive dilepton invariant mass spectra with expansion-sensitive photon transverse momentum spectra to constrain the system's radial expansion.}
However, such methods provide only integrated information and are sensitive to nonflow effects such as resonance decays and jet fragmentation, failing to provide a differential view of the underlying fluctuations or collectivity associated with isotropic expansion.

To address these limitations, a new observable, $v_0(p_T)$, has been proposed as a $p_T$-differential measure of radial flow fluctuations~\cite{Gardim:2019iah,Schenke:2020uqq, Parida:2024ckk}. Defined as a normalized covariance between the particle yield in a given $p_T$ bin and the event-wise fluctuation of the mean $p_T$, and evaluated using pseudorapidity-separated subevents, $v_0(p_T)$ suppresses short-range nonflow correlations while isolating long-range collective effects. Theoretical studies using hydrodynamic simulations show that $v_0(p_T)$ exhibits a negative-to-positive sign change with increasing $p_T$, as well as a mass ordering among particle species at low $p_T$, both of which are signatures of collective expansion~\cite{Schenke:2020uqq,Parida:2024ckk}.

Building upon these theoretical developments, the ALICE and ATLAS Collaborations have recently presented the first experimental measurements of $v_0(p_T)$ in Pb+Pb collisions at $\snn = 5.02$ TeV \cite{ATLAS:2025ztg,ALICE:2025iud}. ALICE measured $v_0(p_T)$ for inclusive charged hadrons as well as identified pions, kaons, and protons~\cite{ALICE:2025iud}. The data confirmed several hydrodynamic predictions, including a characteristic mass hierarchy. In parallel, ATLAS reported measurements for charged particles and explored the scaling behavior of the dimensionless ratio $v_0(p_T)/v_0$, suggesting it may exhibit approximate universality with respect to system size and centrality~\cite{Schenke:2020uqq,Parida:2024ckk,ATLAS:2025ztg}. These initial experimental results, alongside comparisons with state-of-the-art hydrodynamic models, have already demonstrated that $v_0(p_T)$ is highly sensitive to the bulk viscosity of the QGP \cite{ATLAS:2025ztg,ALICE:2025iud,Parida:2024ckk} and its equation of state (EoS) \cite{Gong:2024lhq,ALICE:2025iud}.\footnote{A recent study using a simple blast-wave model also shows that the shape of $v_0(p_T)$ is sensitive to both freeze-out parameters and their fluctuations, such as the freeze-out temperature and radial flow velocity \cite{Saha:2025nyu}.}

Despite these important advancements, a comprehensive study of $v_0(p_T)$ across a broad range of beam energies is still lacking.\footnote{It is noted, however, that a study for the Beam Energy Scan energies appeared recently during the preparation of this draft \cite{Jahan:2025cbp}.}
Whether the same features and scaling behaviors persist at lower energies is an open question. This work extends the investigation of $v_0(p_T)$ to the LHC energy $\snn = 2.76$ TeV in Pb+Pb collisions and the top RHIC energy of $\snn = 200$ GeV in Au+Au collisions to examine whether the observed signatures of radial flow are universal features of QGP evolution across different collision energies. Of particular interest is whether the rescaled observable $v_0(p_T)/v_0$ retains its near-universal behavior, potentially providing a new energy-independent probe of the system’s collective dynamics.

The remainder of this paper is organized as follows. Sec.~\ref{sec:model} describes the model and computational framework used for the multistage hydrodynamic simulations. Sec.~\ref{sec:results} presents a systematic investigation of the sensitivity of $v_0(p_T)$ to bulk viscosity, particlization off-equilibrium corrections, hadronic rescattering, and initial-state granularity, along with a discussion of the universal scaling behavior. Finally, Sec.~\ref{sec:conclusion} summarizes the key findings and their implications for constraining the properties of the QGP.

\section{Model and framework}\label{sec:model}

\subsection{Multistage hydrodynamic model}
This work employs the state-of-the-art multistage hydrodynamic model~\cite{JETSCAPE:2020shq, JETSCAPE:2020mzn,JETSCAPE:2022cob} implemented in the JETSCAPE framework (v3.7) \cite{Putschke:2019yrg} to simulate relativistic heavy-ion collisions. This modular framework integrates several essential components describing different stages of the bulk evolution, modeled by \trento+free-streaming+\music+\iss+\SMASH. 

Initial conditions are modeled using the \trento{} model~\cite{Moreland:2014oya}, which describes energy deposition in the transverse plane. This incorporates event-by-event fluctuations in nucleon positions and entropy deposition, characterized by the generalized mean parameter $p$. A brief period of pre-equilibrium dynamics approximates the early out-of-equilibrium evolution with free-streaming \cite{Broniowski:2008qk,Liu:2015nwa}, followed by Landau matching to hydrodynamic variables at a switching time $\tau_{\mathrm{fs}}$. The hydrodynamic evolution of the QGP is then simulated using second-order viscous hydrodynamics (\music)~\cite{Schenke:2010nt,Schenke:2011bn,Paquet:2015lta}. This evolution utilizes a lattice-based EoS \cite{HotQCD:2014kol} and includes temperature-dependent specific shear and bulk viscosities, $\eta/s(T)$ and $\zeta/s(T)$ \cite{Bernhard:2019bmu,JETSCAPE:2020mzn}. Finally, the system undergoes particlization and hadronic transport upon cooling to a switching temperature $T_{\mathrm{sw}}$. At this point, it transitions to a microscopic description of the hadronic phase using the \SMASH{} transport model~\cite{SMASH:2016zqf}. The hadrons are sampled on the freezeout surface using the Cooper-Frye prescription \cite{Cooper:1974mv} via the \iss{} sampling module \cite{Shen:2014vra}, and various schemes for viscous corrections to the distribution function \cite{McNelis:2019auj}, including Grad's and Chapman-Enskog models, are considered.

The model used in this study was calibrated via Bayesian inference applied to a broad set of $p_T$-integrated hadronic observables from both RHIC Au+Au collisions at $\snn=200$ GeV and LHC Pb+Pb collisions at $\snn=2.76$ TeV experiments, respectively. This calibration, performed in a prior study by the JETSCAPE Collaboration~\cite{JETSCAPE:2020shq, JETSCAPE:2020mzn,JETSCAPE:2022cob}, focused on simulating collisions in (2+1)-dimensions under the assumption of boost invariance. The data used in the inference included identified particle yields and mean transverse momentum $\langle p_T \rangle$, charged hadron multiplicity $dN_{\mathrm{ch}}/d\eta$, transverse energy $dE_T/d\eta$, mean-$p_T$ fluctuations $\delta p_T/\langle p_T \rangle$, and anisotropic flow coefficients $v_n\{2\}$ for $n=2,3,4$.

The Bayesian analysis revealed that while the data strongly constrain $\eta/s(T)$ and $\zeta/s(T)$ in the deconfined crossover region ($T \sim 150{-}250 \, \mathrm{MeV}$), the viscosities at higher temperatures ($T \gtrsim 250 \, \mathrm{MeV}$) remain poorly constrained, with posterior distributions largely reflecting prior assumptions. Notably, the extracted bulk viscosity shows considerable model dependence and wider uncertainty than the shear viscosity. The Grad's viscous correction model is favored by Bayes' evidence, while the Chapman-Enskog model is strongly disfavored due to its difficulty in simultaneously describing pion and proton yields \cite{JETSCAPE:2020shq}.

This study uses the Maximum A Posteriori (MAP) parameter set obtained in~\cite{JETSCAPE:2020mzn} to perform forward model calculations and investigate the new observable $v_0(p_T)$ \cite{Gardim:2019iah,Schenke:2020uqq}. This new observable is believed to have potential sensitivity to bulk viscous effects and may offer complementary constraints to those provided by conventional soft hadronic measurements. By comparing model predictions for the novel observable against recent experimental measurements, this study aims to test the predictive power of the calibrated model and explore its utility in further constraining the QGP transport properties, particularly the bulk viscosity.

\subsection{Model performance at MAP parameters}

\begin{figure}[!t]
  \centering
  \includegraphics[width=0.82\linewidth]{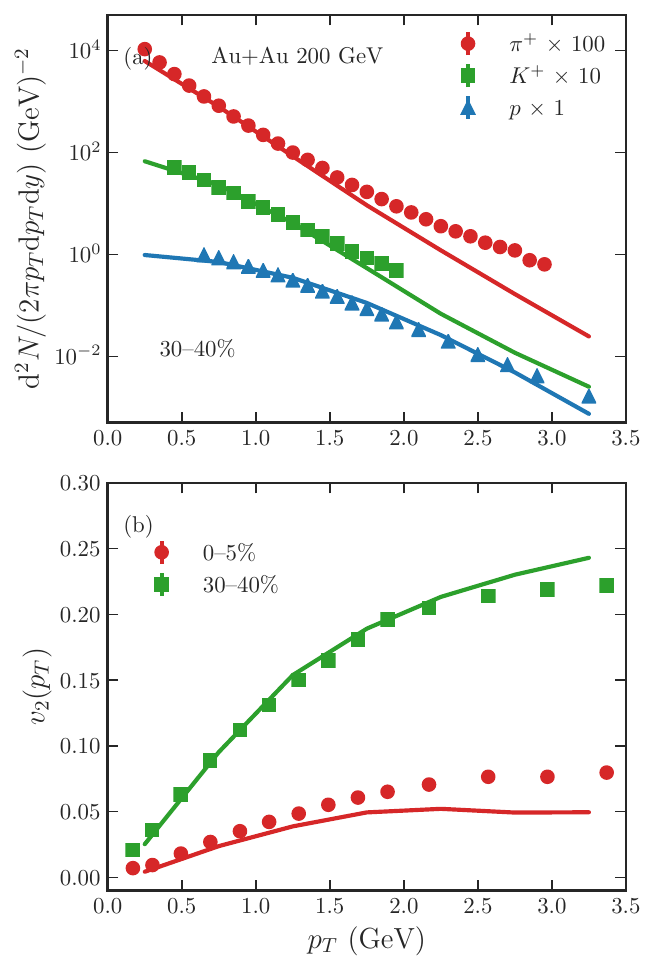}
    \caption{Validation of the multistage hydrodynamic model with MAP parameters, against experimental data from Au+Au collisions at $\snn = 200$~GeV \cite{PHENIX:2003iij,STAR:2004jwm}. (a) Transverse momentum spectra of identified hadrons ($\pi^+$, $K^+$, $p$) in the 30--40\% centrality class. (b) Elliptic flow $v_2(p_T)$ of charged hadrons for two centrality classes (0--5\% and 30--40\%).
    }
  \label{fig:map_ptdiff}
\end{figure}

To validate the quality of the Bayesian-calibrated model used in this study, a comparison between simulation results obtained with the Maximum A Posteriori (MAP) parameter set and experimental data is made. While the model was originally calibrated using $p_T$-integrated observables from both RHIC and LHC experiments, it is shown that it successfully reproduces key $p_T$-differential distributions that were not part of the Bayesian inference, thereby providing further confidence in its predictive power. 

Specifically, Figure~\ref{fig:map_ptdiff} displays the transverse momentum spectra of identified hadrons ($\pi^+$, $K^+$, $p$) in 30--40\% centrality and the $v_2(p_T)$ of charged hadrons in the 0--5\% and 30--40\% centrality classes at 200 GeV \cite{PHENIX:2003iij,STAR:2004jwm}. These observables are particularly sensitive to the hydrodynamic evolution and the temperature dependence of transport coefficients. This predictivity test at RHIC energies complements a similar validation performed at the LHC energy of 2.76 TeV by the JETSCAPE collaboration in Ref.~\cite{JETSCAPE:2020mzn}, with the resulting agreements and discrepancies showing similar features.\footnote{Similar validation between model and data for $p_T$-integrated observables is also confirmed by the simulations but not shown.}

As shown in Fig.~\ref{fig:map_ptdiff}, the agreement between model predictions and data is good at low $p_T$, where the bulk of particle production occurs and the system is well-described by hydrodynamics. However, at higher $p_T$, a consistent discrepancy emerges, with the measured spectra consistently lying above the model calculations. This suggests that while the model successfully captures the collective dynamics of the bulk medium, it underpredicts the production of high-$p_T$ hadrons. This may indicate that the model is either missing key physics, such as a more detailed description of jet-medium interactions or non-equilibrium dynamics, or that the calibration was primarily sensitive to low-$p_T$ observables. The following sections demonstrate how the new observable $v_0(p_T)$ provides complementary information to probe these missing physics aspects.

\section{Results and discussion}\label{sec:results}

Having validated the calibrated model against a range of experimental data not used in the original inference, it is now applied to investigate the radial flow coefficient, $v_0(p_T)$. The global radial flow observable, $v_0$, quantifies the event-by-event fluctuations in the mean transverse momentum \cite{Schenke:2020uqq}, denoted as $[p_T]$. It is defined as $v_0 = \sigma_{[p_T]}/\langle [p_T] \rangle$, where $\sigma_{[p_T]}$ is the standard deviation of the event-wise mean-$p_T$.

The observable $v_0(p_T)$ is a $p_T$-differential measure of radial flow fluctuations. It is defined as a normalized covariance between the event-wise particle yield in a given $p_T$ bin ($\delta n(p_T)$) and the event-wise fluctuation of the mean transverse momentum ($\delta [p_T]$) \cite{Schenke:2020uqq}. The general expression is given by:
\begin{equation}\label{eq:def}
v_0(p_T) = \frac{\langle \delta n(p_T) \cdot \delta [p_T] \rangle}{\langle n(p_T) \rangle \cdot \sigma_{[p_T]}}\,,
\end{equation}
where $n(p_T)$ is the normalized event-by-event particle yield in a given $p_T$ bin, $[p_T]$ is the mean transverse momentum in the event, $\delta x = x - \langle x \rangle$ denotes fluctuations from the ensemble average, and $\sigma_{[p_T]}$ is the standard deviation of $[p_T]$ across events.

The definition in Eq.~\eqref{eq:def} is analogous to a Pearson correlation coefficient \cite{Schenke:2020uqq}. Physically, $v_0(p_T)$ reflects how fluctuations in radial flow strength (captured by $[p_T]$) reshape the single-particle spectrum $n(p_T)$. In hydrodynamic models, increased radial flow boosts particles to higher $p_T$, flattening the spectrum. As a result, $v_0(p_T)$ is negative at low $p_T$ and positive at high $p_T$, with a characteristic zero-crossing that typically occurs near the mean transverse momentum of the event \cite{Schenke:2020uqq, Parida:2024ckk}.

\subsection{Impact of method and kinematic cuts}\label{sec:method}

To compute $v_0(p_T)$ defined in Eq.~\eqref{eq:def} and to compare it with experimental measurements, two complementary formulations are used: a covariance-based approach and a fluctuation-based approach. 
Both methods are constructed to minimize short-range non-flow correlations by correlating the transverse momentum distribution in subevent~A with the mean transverse momentum in subevent~B, separated by a pseudorapidity gap. 
These two formulations, corresponding to the experimental practices of ALICE and ATLAS, respectively, are introduced below.

\ld{Two equivalent formulations of \(v_0(p_T)\) have been adopted in experimental analyses.  
The first, used by ALICE~\cite{ALICE:2025iud}, expresses \(v_0(p_T)\) in terms of ensemble-averaged quantities,
\begin{equation}\label{eq:cov}
v_0^{\text{(cov)}}(p_T) =
\frac{\langle n_A(p_T)\,[p_T]_B\rangle
- \langle n_A(p_T)\rangle\langle [p_T]_B\rangle}
{\langle n_A(p_T)\rangle\,\sigma_{[p_T]}}\,,
\end{equation}
which is referred to as the \emph{direct covariance form}.  
The second, following the ATLAS analysis~\cite{ATLAS:2025ztg}, is expressed in terms of event-by-event fluctuations,
\begin{equation}\label{eq:fluct}
v_0^{\text{(fluct)}}(p_T)
= \frac{\langle \delta n(p_T)\,\delta [p_T]\rangle}
{\langle n(p_T)\rangle\,\sigma_{[p_T]}}\,,
\end{equation}
and will be called the \emph{symmetrized fluctuation form}.  
Formally, the two expressions are identical, since the covariance is defined as the ensemble average of the product of fluctuations.  
In practice, however, they may differ slightly depending on the specific symmetrization and subevent definitions employed in the measurement.  
In the present study, the main difference between the two forms lies in their symmetrization procedure, which is detailed in Appendix~\ref{app:v0_calc}.\footnote{Depending on the analysis setup, the subevent indices may appear explicitly in either formulation; see Appendix~\ref{app:v0_calc} for the corresponding symmetrized definitions.}
}

Results from both methods are compared to assess their consistency and robustness. Additionally, both methods are implemented using two different kinematic settings in $p_T$.\footnote{In this study, the same $p_T$ cuts are applied to both subevents A and B for all model calculations, whereas in some experimental analyses (e.g., at ATLAS \cite{ATLAS:2025ztg}), different kinematic cuts may be used for each subevent to optimize event selection or reduce non-flow effects.} One setting has no low-$p_T$ cutoff, including particles with $p_T \to 0$ as is possible in theoretical simulations. The other applies a low-$p_T$ cutoff consistent with experimental acceptance: $p_T > 0.2$ GeV for ALICE \cite{ALICE:2025iud} and $p_T > 0.5$ GeV for ATLAS \cite{ATLAS:2025ztg}. \ld{All results are obtained on an event-by-event basis using a sufficient number
of hydrodynamic events per model configuration, with statistical uncertainties estimated through a jackknife-like procedure; details are provided in Appendix~\ref{app:events}.}

\begin{figure}[!t]
    \centering
    \includegraphics[width=0.85\linewidth]{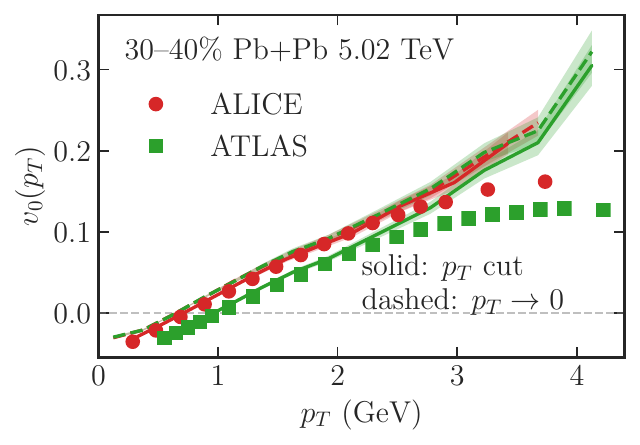}
    \caption{%
    Comparison of model results for $v_0(p_T)$ with ALICE \cite{ALICE:2025iud} and ATLAS \cite{ATLAS:2025ztg} measurements for charged hadrons in 30--40\% Pb+Pb collisions at 5.02~TeV. Solid lines represent model results with low-$p_T$ cuts matching experimental acceptance, while dashed lines show results with no low-$p_T$ cut applied. For the no-cut case, the model results corresponding to both measurements are largely overlapping.
    }
    \label{fig:cut_comparison}
\end{figure}

To evaluate the phenomenological relevance of the model results for $v_0(p_T)$, the results are compared with experimental measurements from ALICE and ATLAS at $\snn = 5.02$ TeV in the 30--40\% centrality class. Figure~\ref{fig:cut_comparison} compares the model results\footnote{The curves exhibit some statistical fluctuations (``wiggles''), particularly at high $p_T$, where larger $p_T$ bins are implemented. A more detailed discussion of the physical origin of these fluctuations will be presented in a later section.} for charged hadrons using the fluctuation-based method to experimental data from ALICE and ATLAS. The solid lines represent the model's predictions with $p_T$ cuts that match the corresponding experimental measurements, while the dashed lines show the model's behavior when the low-$p_T$ cut is removed ($p_T \to 0$).

One first observes that at low transverse momentum, the model results (solid lines) show good agreement with the experimental data. However, this agreement holds up to a different $p_T$ value for each experiment, extending to approximately $p_T \lesssim 2.5$ GeV for ALICE and $p_T \lesssim 2.2$ GeV for ATLAS. Beyond this range, a clear discrepancy emerges: the experimental values for $v_0(p_T)$ flatten out or bend downwards (when going to even higher $p_T$ regions beyond shown in the figure), whereas the model predictions continue to rise with increasing slopes. Both the horizontal difference between the ALICE and ATLAS measurements and the different $p_T$ values where the model-data discrepancy becomes apparent can be attributed to the distinct kinematic cuts used by the experiments. Specifically, the lower $p_T$ cut used by ALICE, resulting in a smaller mean $p_T$, shifts the overall curve horizontally compared to the ATLAS measurement with a higher low-$p_T$ cut.

This effect is further highlighted when the low-$p_T$ cut is removed entirely. In this case, the model results for ALICE and ATLAS (dashed lines) almost completely overlap, as they now consider the same particle population (identical in $p_T$ range, though not necessarily in rapidity). This provides strong evidence that the differences between the two datasets are largely a result of their distinct experimental acceptances. The dashed curves are shifted leftward compared to the solid lines, reflecting the smaller mean $p_T$ of the particle ensemble when very low-momentum particles are included (see also Refs.~\cite{Parida:2024ckk,Bhatta:2025oyp}). Consequently, the green dashed curve is now significantly displaced from the corresponding ATLAS measurement, which was obtained with a higher low-$p_T$ cut. This result highlights the importance of matching experimental acceptance in simulations; it also shows that while doing so is essential for comparing model results to different datasets, it does not resolve the discrepancy seen at high $p_T$.

\begin{figure}[!t]
    \centering
    \includegraphics[width=0.83\linewidth]{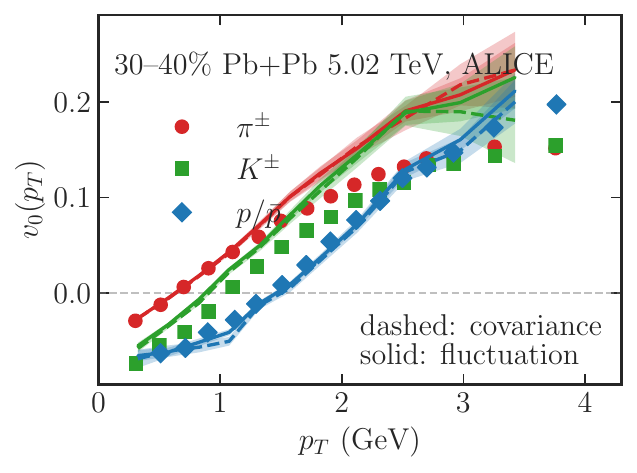}
    \caption{%
    The $p_T$-differential $v_0(p_T)$ for identified hadrons ($\pi^\pm$, $K^\pm$, $p/\bar{p}$) in 30--40\% Pb+Pb collisions at 5.02~TeV from ALICE \cite{ALICE:2025iud}. Solid lines show results from the fluctuation-based method, while dashed lines show the covariance-based method.
    }
    \label{fig:method_species}
\end{figure}

Extending this analysis to identified particles in Fig.~\ref{fig:method_species}, the model successfully reproduces the mass ordering of $v_0(p_T)$ at low transverse momentum, a key signature of collective radial flow. However, the high-$p_T$ discrepancy observed for charged hadrons persists for all identified species, where the model predictions continue to rise while the data flatten out. The $p_T$ value where this discrepancy begins is mass-dependent, appearing at a lower $p_T$ for lighter particles. It is also notable that the model consistently overpredicts the $v_0(p_T)$ of kaons across the entire measured $p_T$ range; a similar discrepancy for kaons can be seen in Ref.~\cite{ALICE:2025iud}. \ld{
A more detailed investigation of this species-dependent deviation would be valuable 
but lies beyond the scope of the present study.}

\begin{figure*}[!t]
    \centering
    \includegraphics[width=0.92\linewidth]{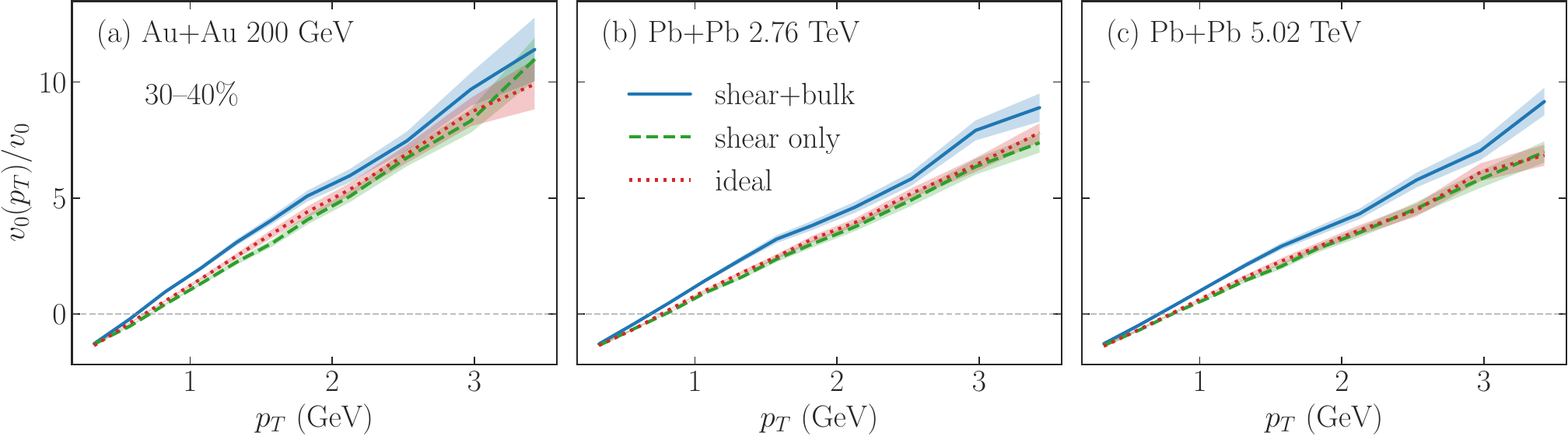}\\
    \vspace{4pt}
    \includegraphics[width=0.92\linewidth]{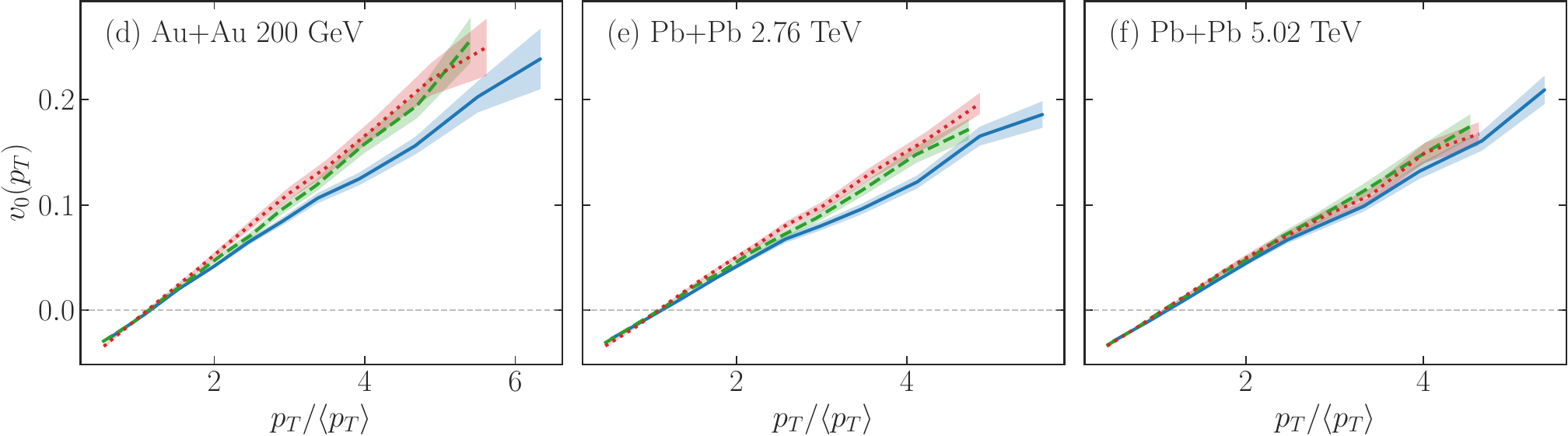}
    \caption{%
        Upper panels: Transverse momentum dependence of the scaled radial flow $v_0(p_T)/v_0$ for charged hadrons at three beam energies: (a) $\snn = 200$~GeV, (b) $\snn = 2.76$~TeV, and (c) $\snn = 5.02$~TeV, all for the 30--40\% centrality class, as a function of $p_T$. Lower panels: Unscaled $v_0(p_T)$ as a function of the rescaled variable $p_T/\langle p_T \rangle$. Results are shown for three scenarios: full viscous evolution with shear and bulk viscosity (solid blue), shear-only viscosity (dashed green), and ideal hydrodynamics (dotted red).
    }
    \label{fig:bulk_viscosity}
\end{figure*}

The high-$p_T$ discrepancy for $v_0(p_T)$ is closely related to the model's underprediction of the transverse momentum spectra of identified hadrons at high $p_T$ shown in Fig.~\ref{fig:map_ptdiff}(a). However, while they might originate from the same missing physics, the $v_0(p_T)$ observable provides complementary information to the spectra. This is because $v_0(p_T)$ directly probes the correlation between a particle's momentum and the event-wise collective flow, information that is not fully captured by the spectra alone. For example, even if the model could be adjusted to produce more high-$p_T$ particles to match the spectra, it would likely still fail to reproduce the flattening of $v_0(p_T)$, possibly as a signature of weakened coupling to collective fluctuations.\footnote{Unfortunately, the current model is not flexible enough to be re-calibrated to produce a matched high-$p_T$ spectrum, which would be necessary to precisely quantify the degree of flattening in $v_0(p_T)$. This limitation prevents us from directly estimating the missing physics, such as a decoupled high-$p_T$ particle source, that weakens the collective fluctuations. A dedicated analysis using the framework of Ref.~\cite{Jia:2025rab}, which disentangles kinematic effects from genuine dynamical effects via the $g(p_T)$ function, could provide further insight into this discrepancy.}

Therefore, the common failure to describe both the high-$p_T$ spectra and $v_0(p_T)$ for identified species points to the same underlying deficiency in the model: the insufficient description of a source of high-$p_T$ particles that are decoupled from the collective expansion. Incorporating such a source would not only increase the high-$p_T$ yield to match the spectra but would also dilute the overall covariance (i.e., the numerator of Eq.~\eqref{eq:def}), causing the $v_0(p_T)$ to flatten out as observed in the data. The consistent overprediction of kaon $v_0(p_T)$ across the entire $p_T$ range, while good agreement is found for low-$p_T$ spectra, might further suggest a potential issue with the model's description of strange hadron coupling to the collective dynamics. This highlights the utility of $v_0(p_T)$ as a stringent test of collective dynamics, offering discriminating power for models that is more sensitive to these effects than the spectra alone.

\subsection{Sensitivity to bulk viscosity}\label{sec:bulk}

Previous studies at $\snn=5.02$ TeV have identified $v_0(p_T)$, normalized by $v_0$, as a sensitive probe of bulk viscosity in relativistic heavy-ion collisions~\cite{Parida:2024ckk,ALICE:2025iud,ATLAS:2025ztg}. These studies observed that turning on bulk viscosity notably steepens the slope of $v_0(p_T)/v_0$ as a function of $p_T$, while shear viscosity has a comparatively minor effect. In this section, this finding is tested using the model described in Sec.~\ref{sec:model}, which is distinct from previous ones, to check the robustness of the observation across different modeling approaches. Additionally,  this analysis is extended to lower beam energies, such as $\snn = 2.76$~TeV and the top RHIC energy of $\snn=200$~GeV. Here, this study focuses on $v_0(p_T)$ as a probe of the underlying collectivity of the bulk medium rather than comparing to experiments, so particles down to $p_T=0.2$ GeV are included, which are an essential part of the collective flow.

For this purpose, three scenarios are investigated: a default case with both shear and bulk viscosities (``shear+bulk''), a case with bulk viscosity off (``shear only''), and a case with both bulk and shear viscosities off (``ideal''). The comparisons in Fig.~\ref{fig:bulk_viscosity}(c) for $\snn = 5.02$~TeV confirm the previously observed trend: the bulk-viscous case exhibits a visibly stronger rise in $v_0(p_T)/v_0$ compared to both the shear-only and ideal cases, while the latter two cases show only minor differences. This sensitivity arises from the role of bulk pressure in damping radial expansion during the hydrodynamic evolution \cite{Noronha-Hostler:2013gga,Ryu:2015vwa}, thereby reducing the transverse momentum imparted to particles and shifting the curve of $v_0(p_T)/v_0$ to the left, i.e., to smaller $p_T$ regions.

However, when extending this analysis to lower beam energies, from $\snn = 2.76$~TeV down to $\snn = 200$~GeV, the distinction between the shear+bulk and the other two cases becomes less pronounced (see Fig.~\ref{fig:bulk_viscosity}(a) and (b)). The difference between the shear-only and ideal cases becomes more prominent, and the curve for $v_0(p_T)/v_0$ from ideal hydrodynamics lies between those from the bulk+shear and shear-only cases. This intermediate behavior suggests a more nuanced interplay: while bulk viscosity isotropically reduces pressure and thus suppresses radial expansion, shear viscosity can enhance early-time transverse pressure gradients by anisotropically redistributing momentum, potentially increasing $\langle p_T \rangle$ \cite{Song:2009rh} and thus reducing the slope of $v_0(p_T)/v_0$ compared to the ideal case. This ``crossover'' in behavior with decreasing beam energy is a subtle but important finding, as it suggests that the interplay between shear and bulk viscosity changes with collision energy. At lower energies where the medium is shorter-lived, the relative contributions of these two damping mechanisms become more complex.

Since scaling the vertical axis by $v_0$, i.e., plotting $v_0(p_T)/v_0$ versus $p_T$, could not clearly separate the effects of bulk viscosity at 200 GeV, another rescaling strategy is explored aimed at disentangling these effects and isolating the underlying dynamics of transverse expansion: scaling the horizontal axis by the mean transverse momentum $\langle p_T \rangle$, i.e., plotting $v_0(p_T)$ versus $p_T/\langle p_T \rangle$ (see the second row of Fig.~\ref{fig:bulk_viscosity}). These two approaches emphasize different aspects of the observable. The first form, $v_0(p_T)/v_0$, highlights the shape of the $p_T$ dependence relative to the integrated flow $v_0$, but can obscure effects that suppress or strengthen both the numerator and denominator in the same way. In contrast, the second form, $v_0(p_T)$ as a function of $p_T/\langle p_T \rangle$, explicitly normalizes out the blue-shift associated with varying radial flow (indicated by $\langle p_T \rangle$) and provides a clearer view of how flow fluctuations are distributed across the spectrum.

\begin{figure*}[!t]
    \centering
    \includegraphics[width=0.92\linewidth]{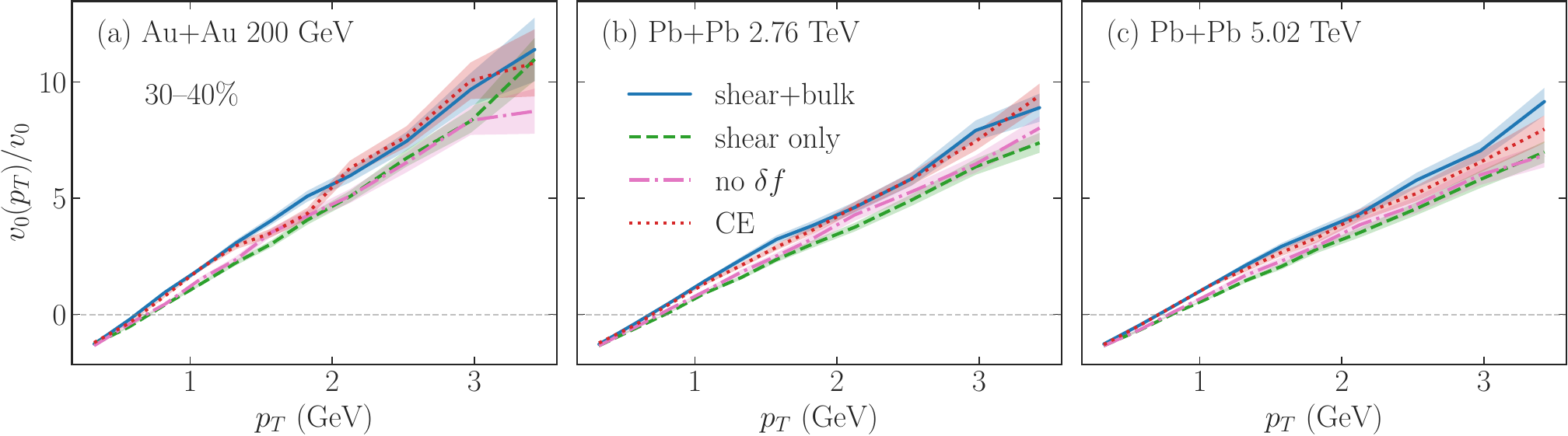}
    \caption{
    Scaled radial flow observable $v_0(p_T)/v_0$ for charged hadrons in the 30--40\% centrality class at three collision energies: (a) $\snn = 200$~GeV, (b) 2.76~TeV, and (c) 5.02~TeV. Four model scenarios are compared: the default case with both shear and bulk viscosity and Grad viscous corrections (``shear+bulk'', solid blue), shear viscosity only (``shear only'', dashed green), no viscous corrections to the distribution function (``no~$\delta f$'', dot-dashed megenta), and Chapman--Enskog viscous corrections (``CE'', dotted red).
    }
    \label{fig:off_eq}
\end{figure*}

This second approach proves particularly informative at $\snn = 200$~GeV. The lower panels of Figure~\ref{fig:bulk_viscosity} show that when plotted against $p_T/\langle p_T \rangle$, the bulk+shear curve shifts rightward relative to the shear-only and ideal cases. This reflects the fact that bulk viscosity significantly reduces $\langle p_T \rangle$, thereby magnifying differences in the normalized spectral shape. In other words, the bulk viscosity's reduced mean $p_T$ enhances the separation between it and the other model variants in this rescaled frame, and makes such a rescaling method more useful for constraining the bulk viscosity at $\snn = 200$~GeV. This enhanced sensitivity at lower energies arises because bulk viscosity has a more pronounced relative effect on the weaker radial flow and mean transverse momentum at lower collision energies. The normalization by $\langle p_T \rangle$ in this approach effectively ``magnifies'' the differences in the underlying flow dynamics. Nevertheless, such a rescaling method makes the curves of these model variants closer to each other and thus would not help to isolate the effects of bulk viscosity at $\snn = 5.02$ TeV. At 5.02 TeV, where radial flow is much stronger, the relative suppression caused by bulk viscosity is less dramatic, making the normalization less powerful as a discriminating tool.

In conclusion, while $v_0(p_T)/v_0$ remains a useful probe of bulk viscosity, particularly at 5.02 TeV, alternative scalings such as $v_0(p_T)$ versus $p_T/\langle p_T \rangle$ offer enhanced sensitivity to bulk-viscous effects at lower beam energies such as 200 GeV. The fact that the sensitivity of $v_0(p_T)$ to bulk viscosity changes with beam energy suggests that the model is probing how the QGP's properties, including its EoS and transport coefficients, evolve across a wide range of initial conditions. These scaling strategies are not merely decorative; they improve the ability to differentiate between model settings and thus help constrain essential properties of the QGP medium, such as its bulk viscosity and radial expansion, across different initial conditions.

\subsection{Sensitivity to off-equilibrium corrections}

To investigate the effects of off-equilibrium corrections applied during particlization, simulations are performed using the Cooper--Frye prescription with several implementations of viscous corrections to the distribution function. The default viscous correction model is the Grad-type, which is applied to all cases except the ``CE'' setting. In the CE case, the Chapman--Enskog (``CE'') viscous correction model is used, where the hydrodynamic evolution is the same as in the shear+bulk case. In the ``no $\delta f$'' setting, both shear and bulk viscous corrections to the particle momentum distribution are turned off, while keeping the hydrodynamic evolution unchanged. Their hadronic afterburner stages evolve particles sampled by these various methods accordingly.

As shown in Fig.~\ref{fig:off_eq}, at all three beam energies ($\snn = 0.2$, 2.76, and 5.02~TeV), the $v_0(p_T)/v_0$ curves for the ``no $\delta f$'' case lie close to those of the ``shear-only'' scenario, and are separated from the ``shear+bulk'' case. The ``CE'' case, representing an alternative viscous correction, falls between the ``shear+bulk'' and ``no $\delta f$'' cases in terms of slope while being closer to the former. The hierarchy of slopes is consistently observed as ``shear+bulk'' > ``CE'' > ``no $\delta f$'' > ``shear only''. This hierarchy reflects how different $\delta f$ models impact the radial expansion. The ``shear+bulk'' case, with its aggressive suppression of radial expansion due to bulk viscosity, yields the smallest mean transverse momentum and thus the steepest slope. Conversely, the ``shear only'' and ``no $\delta f$'' cases, with their larger $\langle p_T \rangle$, result in less steep curves. The intermediate slope of the ``CE'' model suggests it imposes a less severe suppression of radial expansion compared to the Grad-type corrections used in the ``shear+bulk'' scenario.

In summary, the observable $v_0(p_T)/v_0$ is demonstrably sensitive to the presence, strength, and form of off-equilibrium corrections applied at particlization. The choice between Grad and Chapman--Enskog models introduces measurable differences, especially in the shape and slope of its $p_T$ dependence. Thus, attempting to constrain particlization models should consider including $v_0(p_T)/v_0$, and test across beam energies.

\begin{figure*}[!t]
    \centering
    \includegraphics[width=0.92\linewidth]{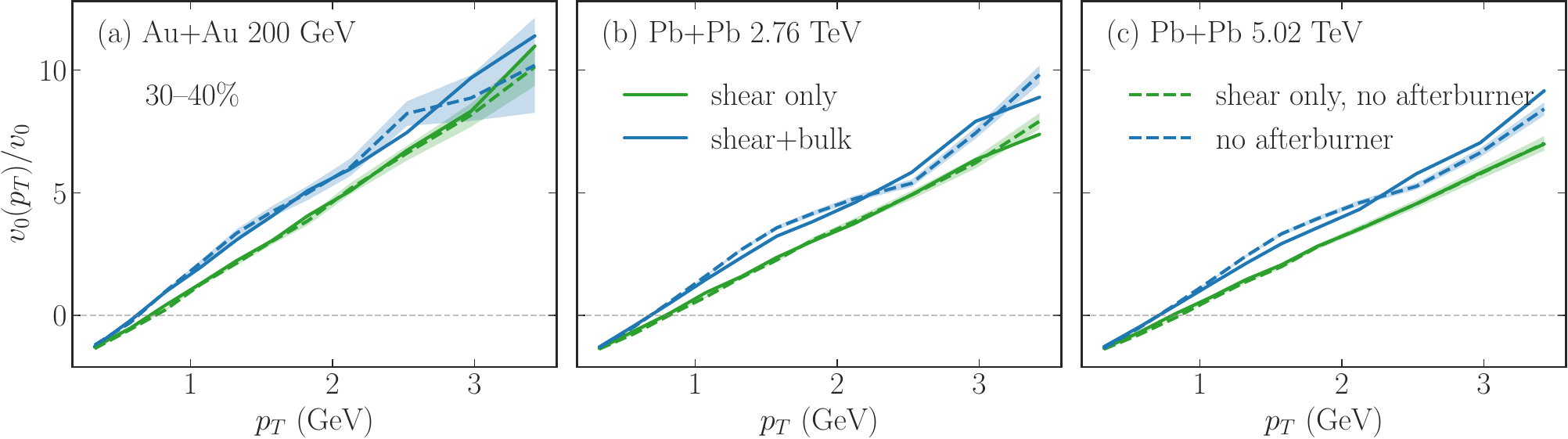}
    \caption{
    Comparison of the scaled radial flow observable $v_0(p_T)/v_0$ with and without the hadronic afterburner for three collision energies: (a) $\snn = 200$~GeV, (b) 2.76~TeV, and (c) 5.02~TeV, all in the 30--40\% centrality class. The simulations consider both ``shear-only'' (green) and ``shear+bulk'' (blue) hydrodynamic scenarios. Solid lines indicate results with the hadronic afterburner, while dashed lines correspond to simulations without it.
    }
    \label{fig:afterburner}
\end{figure*}

\subsection{Sensitivity to hadronic afterburner}

To assess the role of the hadronic phase on $v_0(p_T)$, results obtained with and without the hadronic afterburner are compared, where hadronic rescatterings and resonance decays happen, which further affect the radial expansion. The default simulation (``shear+bulk'') includes full hydrodynamics with both shear and bulk viscosities, followed by a hadronic afterburner. Three reference scenarios are used for comparison: (1) ``no-afterburner'': afterburner turned off with bulk and shear viscous hydrodynamics; (2) ``shear-only'': bulk viscosity turned off, but afterburner on; and (3) ``shear-only, no-afterburner'': both bulk viscosity and afterburner turned off. It is noted that when the afterburner is turned off, the resonance decays are not carried out either, which are considered as part of the hadronic afterburner. To evaluate the effects of the hadronic phase, one should compare ``shear+bulk'' vs. ``no-afterburner'', and ``shear-only'' vs. ``shear-only, no-afterburner'' respectively.

Figure~\ref{fig:afterburner} shows the scaled observable $v_0(p_T)/v_0$ plotted against $p_T$. Across all energies and model settings, one observes that the presence or absence of the hadronic afterburner leads to only minor changes in the shape of $v_0(p_T)/v_0$. The differences between the solid lines (with afterburner) and dashed lines (without afterburner) are noticeably smaller than the differences observed between the ``shear-only'' and ``shear+bulk'' scenarios. Both the ``shear+bulk'' vs. ``no-afterburner'', and ``shear-only'' vs. ``shear-only, no-afterburner'' comparisons indicate that the scaled $v_0(p_T)/v_0$ are relatively insensitive to the dynamics of the late-stage hadronic evolution. This might explain why the model results in Ref.~\cite{Parida:2024ckk} without a hadronic afterburner still agree well with the experimental measurements \cite{ATLAS:2025ztg}.

Nevertheless, one observes stronger statistical fluctuations, or ``wiggles,'' in the curves when the afterburner is off, particularly in scenarios that include bulk viscosity (similar fluctuations are also visible in the bulk-viscous case in Figs.~\ref{fig:bulk_viscosity} and \ref{fig:off_eq}). While a lack of statistics at high $p_T$ certainly contributes to these fluctuations, their prominence in the bulk-viscous scenarios suggests a deeper physical origin. The strong temperature dependence of bulk viscosity, especially near the QCD crossover, along with its non-linear interaction with the fluid's equation of state, can introduce complex features into the momentum distribution. These features may be amplified by the Grad-type viscous corrections, which have a non-linear $p_T$ dependence and are applied to the distribution function at particlization. This interplay makes the observable more susceptible to statistical noise, particularly in the low-yield, high-$p_T$ regime. Conversely, the hadronic afterburner plays a smoothing role, making the radial flow functions more coherent across the $p_T$ spectrum by averaging out some of the complex, local fluctuations that originate during particlization with bulk viscosity.

\begin{figure*}[!t]
    \centering
    \includegraphics[width=0.92\linewidth]{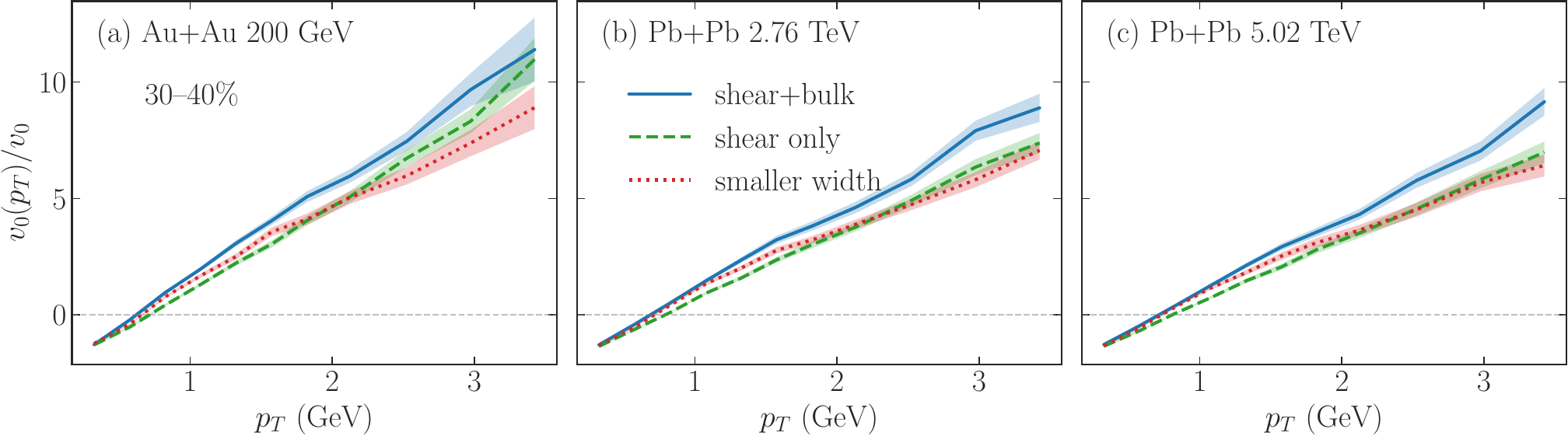}
    \caption{
    Effect of initial-state granularity on the scaled radial flow observable $v_0(p_T)/v_0$, shown for three collision energies: (a) $\snn = 200$~GeV, (b) 2.76~TeV, and (c) 5.02~TeV in the 30--40\% centrality class. Results are compared between the default ``shear+bulk'' (solid blue) and ``shear-only'' (dashed green) scenarios using a nucleon width of $\sigma = 1.12$~fm and a smaller width of $\sigma = 0.8$~fm (dotted red).
    }
    \label{fig:nucleon_size}
\end{figure*}

To explore the effects of the afterburner further, the unscaled version of the observable, $v_0(p_T)$ (not shown), is also examined. It is revealed that turning off the afterburner leads to a moderate change in $v_0(p_T)$. Thus, the effect of the hadronic stage appears more visible in the unscaled observable. This is because $v_0(p_T)/v_0$ normalizes out the overall radial flow strength and emphasizes shape differences, which tend to be minor. The hadronic afterburner's role of re-scattering and resonance decays primarily affects the overall particle yield and mean $p_T$, but it does not significantly alter $v_0(p_T)/v_0$ established during the hydrodynamic phase.

Taken together, these observations suggest that the hadronic afterburner plays a subleading role in shaping the $p_T$-differential radial flow fluctuations. While it modestly changes $v_0(p_T)$, the scaled observable $v_0(p_T)/v_0$ as a function of $p_T$ remains nearly invariant. This underscores the advantage of using $v_0(p_T)/v_0$ to probe early-time dynamics, as it is less contaminated by late-stage rescatterings. Nevertheless, including unscaled observables such as $v_0(p_T)$ provides complementary insights and is valuable when aiming to disentangle effects from different evolution stages.

\subsection{Sensitivity to nucleon width}

So far, this study has fixed the initial condition and investigated the effects of things that affect transverse expansion and their imprint on the radial flow. Now the impact of initial-state granularity is investigated, modeled via the effective nucleon width, on $v_0(p_T)$. Simulations using the default nucleon width from the MAP parameter set ($w = 1.12$~fm), which is used in both ``shear+bulk'' and ``shear-only'' scenarios, are compared to a reduced size of $w = 0.8$~fm, labeled as the ``smaller width'' case. All other model parameters, including transport coefficients and freeze-out prescriptions, are held fixed. The goal is to isolate how changing the transverse smearing of nucleons in the initial conditions affects final-state collective behavior.

As shown in Fig.~\ref{fig:nucleon_size}, decreasing the nucleon width significantly alters the shape of $v_0(p_T)/v_0$. Specifically, the rescaled observable exhibits a downward bending at higher $p_T$ across all beam energies ($\snn=200$~GeV, 2.76~TeV, and 5.02~TeV). Crucially, this downward bending is a robust feature that persists even when varying the bin width of the transverse momentum, indicating that it is a genuine physical effect of the initial-state granularity and not a statistical artifact or a ``wiggle'' in the curve. This deformation is distinct from that induced by bulk viscosity or particlization corrections. 

The bending suggests that enhanced spatial fluctuations from smaller nucleons disrupt the coherence of radial flow across the transverse momentum $p_T$ and thus change the curves' curvature. This shows that reducing the size enhances local fluctuations in the initial energy density profile, which can disrupt the buildup of coherent radial flow, resulting in a distinguishable imprint of initial granularity on the final-state observables. The small-width case leads to a more granular initial state, which produces more localized pressure gradients that are less effective at generating a large-scale, coherent radial flow to boost high-$p_T$ particles in a correlated manner.

It is intriguing to note a qualitative similarity between the model's predictions and experimental observations: the downward bending of the $v_0(p_T)/v_0$ curve from the smaller nucleon width scenario resembles the flattening of the high-$p_T$ experimental data shown in Figs.~\ref{fig:cut_comparison} and \ref{fig:method_species}. This suggests that the mechanism of enhanced initial-state granularity, which disrupts coherent radial flow, could be a contributing factor to the observed high-$p_T$ behavior. However, one must caution that the specific $p_T$ location of the bending in the smaller width model does not precisely align with the high-$p_T$ flattening seen in the data, indicating that this mechanism may be a partial, rather than a complete, explanation.

The sensitivity of $v_0(p_T)$ to nucleon width provides an important lever arm for constraining initial condition models. The downward bending of $v_0(p_T)/v_0$ at high $p_T$ is a distinct feature of the smaller width scenario that is large enough to be discernible above typical experimental uncertainties. Thus, it can serve as a powerful discriminator between different initial-state scenarios. Furthermore, since nucleon width affects $v_0(p_T)$ differently than bulk viscosity or off-equilibrium corrections---where bulk viscosity primarily alters the overall slope and magnitude of the curve, while nucleon width changes its curvature---combining these observables in a multi-dimensional analysis could help disentangle their contributions. This study thus advocates for the inclusion of $v_0(p_T)$ to constrain nucleon width when it is a tunable parameter in future Bayesian model-to-data comparisons.

\subsection{Universality under double rescaling}\label{sec:double}

\begin{figure}[!tb]
    \centering
    \includegraphics[width=0.8\linewidth]{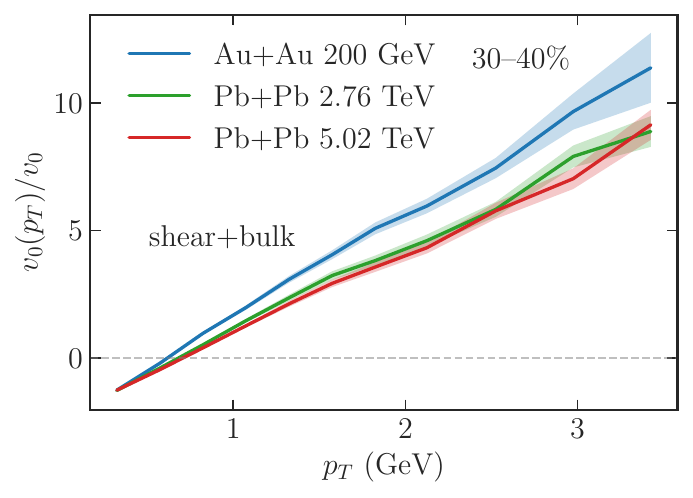}
    \caption{Scaled radial flow fluctuations $v_0(p_T)/v_0$ for charged hadrons in the 30--40\% centrality class from the default ``shear+bulk'' simulation, shown at three beam energies: $\snn = 200$~GeV (blue), 2.76~TeV (green), and 5.02~TeV (red).
    }
    \label{fig:beam_energy}
\end{figure}

To explore the energy dependence of $v_0(p_T)/v_0$ straightforwardly, this study compares the ``shear+bulk'' scenario across the three beam energies, as shown in Fig.~\ref{fig:beam_energy}. The curves for $v_0(p_T)/v_0$ do not overlap; the slope is steepest for the lowest energy (200 GeV), followed by the intermediate energy (2.76 TeV), and is least steep for the highest energy (5.02 TeV).\footnote{This contrasts with observations at even lower Beam Energy Scan energies for the unnormalized $v_0(p_T)$, which shows a smaller slope with decreasing collision energy due to a reduction in radial flow fluctuations \cite{Jahan:2025cbp}.}
It is important to note, however, that these larger normalized results at lower energies should not be misinterpreted as a larger radial flow. Instead, this behavior is a consequence of smaller radial expansion and a lower mean $p_T$ at these energies. The observed energy dependence highlights that the distribution of flow fluctuations across the transverse momentum spectrum, relative to the overall flow magnitude, is different at each beam energy, indicating that a simple vertical rescaling by $v_0$ is not sufficient to remove all energy dependence and that a more comprehensive rescaling strategy might be required to reveal the underlying universal dynamics of flow fluctuations.

To further isolate the collective features embedded in $v_0(p_T)/v_0$, this study explores a combined rescaling strategy: plotting $v_0(p_T)/v_0$ as a function of $p_T/\langle p_T \rangle$.\footnote{%
For a meaningful comparison, $\langle p_T \rangle$ should be calculated from the same particles and within the same kinematic range (e.g., $p_T$ and rapidity cuts) as those used for the calculation of the flow observable $v_0(p_T)$. This ensures that the scaling factor correctly accounts for the average transverse expansion of the system being studied, allowing for a genuine isolation of the shape of flow fluctuations.}
This approach combines two complementary scalings explored in Sec.~\ref{sec:bulk} that highlight different aspects of the observable. The vertical scaling, $v_0(p_T)/v_0$, removes the overall magnitude of the radial flow fluctuations and emphasizes the spectral shape, which is useful for comparing the relative contributions of each $p_T$ bin to the total signal. The horizontal scaling, $p_T/\langle p_T \rangle$, normalizes out the blue-shift effect associated with the average transverse expansion, aligning the location of spectral features across different systems and thereby testing for a universal response pattern. This combined representation simultaneously removes the overall magnitude of radial flow (via $v_0$) and the blue-shift effect due to the average transverse expansion (via $\langle p_T \rangle$), thereby focusing on the intrinsic structure of flow fluctuations across $p_T$.

\begin{figure}[!t]
    \centering
    \includegraphics[width=0.82\linewidth]{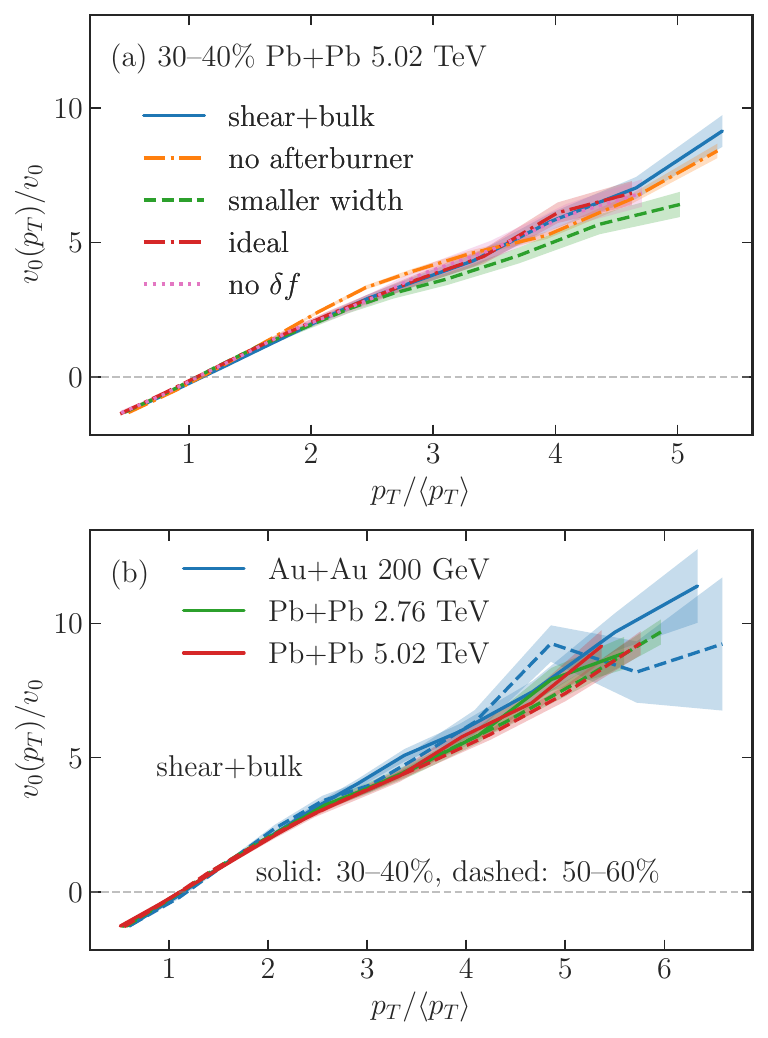}
    \caption{
    Double-rescaled results for $v_0(p_T)/v_0$ as a function of $p_T/\langle p_T \rangle$. (a) Comparisons of different modeling scenarios at 5.02~TeV (30--40\% centrality), including shear+bulk, ideal, no afterburner, and smaller nucleon width cases. (b) Comparison across beam energies and centralities using the default shear+bulk setup.
    }
    \label{fig:double_scaling}
\end{figure}

Figure~\ref{fig:double_scaling} illustrates this behavior in two complementary comparisons. The top panel shows results from multiple model scenarios at a fixed beam energy and centrality, including the default ``shear+bulk'', an ``ideal'' fluid, a ``no afterburner'' case, and scenarios with altered off-equilibrium corrections or nucleon width. The bottom panel focuses on the default ``shear+bulk'' scenario but compares results across multiple beam energies and centrality classes.

In both comparisons, one finds a striking level of universality: all rescaled curves collapse onto a common band for $p_T/\langle p_T \rangle \lesssim 2.5$, with model-specific deviations emerging only at higher values of the rescaled momentum. This behavior is observed not only across beam energies and centrality classes but also across a wide range of physical scenarios, including those with and without bulk viscosity, hadronic afterburners, and off-equilibrium corrections.

\ld{This result reinforces the interpretation that the lower-$p_T$ region ($p_T/\langle p_T \rangle{\,<\,}2.5$) is dominated by collective dynamics that are common across systems and modeling choices. 
The suppression of model-to-model variation in this region indicates that, once the overall scales set by $v_0$ and $\langle p_T \rangle$ are removed, the functional shape of $v_0(p_T/\langle p_T \rangle)/v_0$ is largely universal. 
This universality does not imply that the scaled radial flow is completely insensitive to modeling inputs such as shear or bulk viscosities; rather, their effects are primarily absorbed into the overall normalization and mean momentum scale, while the remaining shape reflects the common hydrodynamic response of the medium. 
In this sense, the observed universality supports the view that the collective expansion follows a universal pattern driven by hydrodynamic flow.

Conversely, the differences that emerge at higher $p_T/\langle p_T \rangle$ reflect model-specific physics that become increasingly important in shaping the tails of the $v_0(p_T)$ distribution. 
For example, the ``ideal'' curve bends upward at high $p_T/\langle p_T \rangle$ due to the absence of viscous damping, while the ``smaller width'' scenario, with its enhanced initial-state granularity, causes a downward bending. 
The ``no afterburner'' and ``no $\delta f$'' cases also diverge, highlighting the importance of particlization and late-stage dynamics in this regime.

In summary, the double-rescaled plot provides a stringent test of collectivity. 
The universality observed at low $p_T/\langle p_T \rangle$ demonstrates that the collective response of the medium exhibits a common hydrodynamic origin across different modeling inputs, while the deviations at higher $p_T$ offer a sensitive window into the underlying transport and freeze-out properties. 
Such scaling behavior provides a robust and practical tool to test the presence of collectivity in new collision systems, including small systems, by examining whether their data collapse onto the same universal band.}

\section{Conclusion}\label{sec:conclusion}

This work has established the transverse momentum differential radial flow observable, $v_0(p_T)$, as a highly sensitive and multifaceted diagnostic tool for probing the collective dynamics of the QGP. By systematically investigating its response to a broad range of physical effects and modeling choices across three collision energies ($\snn = 0.2$, 2.76, and 5.02~TeV), it is demonstrated that distinct features of the observable encode unique information about the entire evolution of the system.

This analysis revealed that the shape and slope of the rescaled observable, $v_0(p_T)/v_0$, are a sensitive probe of transport properties and particlization dynamics. The presence of bulk viscosity, for instance, visibly steepens the curve at LHC energies, a signature that becomes less pronounced at lower beam energies where shear viscosity effects are more dominant. Similarly, the specific form of off-equilibrium corrections ($\delta f$) at particlization introduces measurable changes to the curve's curvature, providing a valuable lever arm to constrain freeze-out models. Crucially, it is found that these dynamics are dominantly established during the hydrodynamic phase, as the observable is largely insensitive to the late-stage hadronic afterburner. Furthermore, the initial-state granularity, modeled by the effective nucleon width, imprints a unique signature in $v_0(p_T)/v_0$ by causing a distinct downward bending at high $p_T$, an effect not seen in any of the other model variations. This makes $v_0(p_T)$ a particularly powerful discriminator between early- and late-stage effects.

An additional layer of insight is gained by applying a double rescaling strategy, plotting $v_0(p_T)/v_0$ as a function of $p_T/\langle p_T \rangle$. This representation reveals a remarkable degree of universality: all studied scenarios, regardless of beam energy, centrality class, viscosity model, or initial granularity, collapse onto a common trend in the low-$p_T$ region ($p_T/\langle p_T \rangle \lesssim 2.5$). This universality strongly suggests that the collective response of the bulk medium is robust and governed by common hydrodynamic principles. Deviations from this universal curve at higher $p_T/\langle p_T \rangle$ then provide a sensitive window into model-specific physics such as viscous damping, particlization details, and initial-state granularity.

Taken together, these findings underscore the importance of carefully matching theoretical calculations to experimental analysis conditions, as illustrated by the substantial impact of low-$p_T$ acceptance cuts. This study advocates for the inclusion of $v_0(p_T)$ observables in future Bayesian inference studies and model-to-data comparisons. The multi-dimensional sensitivity of this observable, revealed through the analysis of its slope, curvature, and dependence on both single and double rescaling, offers an unprecedented opportunity to provide tighter, more robust constraints on the properties of the QGP and its full spacetime evolution.

\section*{Acknowledgements}

The author acknowledges useful conversations with Somadutta Bhatta, Xin Dong, and Ashish Pandav. This work was supported in part by the U.S. Department of Energy, Office of Science, Office of Nuclear Physics under
Grant No.~DE-AC02-05CH11231. Computations were made on the computers managed by the Ohio Supercomputer Center \cite{OhioSupercomputerCenter1987}. The author acknowledges the use of AI-based tools, including Gemini and ChatGPT, for grammar refinement, clarity enhancement, and analysis code optimization. All AI-assisted revisions were carefully reviewed to ensure that the intended meaning and scientific content remained unchanged.

\section*{Data availability}

The data supporting the findings of the model simulation are openly available at Ref.~\cite{du_2025_17808246} on Zenodo.

\appendix*

\appendix
\section{Computational framework for \texorpdfstring{$v_0(p_T)$}{v0(pT)} analysis}\label{app:v0_calc}

To compute the transverse momentum differential observable $v_0(p_T)$, this study implemented a two-stage modular pipeline, mirroring the structure of experimental analysis.

\subsection{Event-level processing}

The first stage is a function that processes event-wise particle lists and computes key intermediate quantities. This function takes as input a particle array (containing PID, $p_T$, and pseudorapidity $\eta$) and applies kinematic cuts on $p_T$ and $\eta$. Particles are then sorted into two subevents, A and B, based on their pseudorapidity. For each subevent and for each particle species (e.g., charged hadrons, pions, etc.), the following quantities are calculated and stored:
\begin{itemize}
    \item The normalized particle yield in each $p_T$ bin, $n_A(p_T)$ and $n_B(p_T)$.
    \item The mean transverse momentum $[p_T]_A$ and $[p_T]_B$.
    \item The product of the normalized yield in one subevent with the mean transverse momentum in the other, i.e., $n_A(p_T) \cdot [p_T]_B$.
\end{itemize}
These per-event results serve as the raw input for the subsequent ensemble-level analysis.

\begin{table*}[!htbp]
\centering
\caption{
Approximate number of simulated hydrodynamic events used in the analysis. 
Values correspond to successfully completed events (rounded to the nearest 500). 
The study primarily focuses on the 30--40\% centrality class at each beam energy (left three columns), 
while results for the 50--60\% class are shown for the ``shear+bulk'' case (rightmost column) for comparison.
}
\begin{tabular}{lcccc}
\hline\hline
Setup & 
Au+Au 200 GeV & 
Pb+Pb 2.76 TeV & 
Pb+Pb 5.02 TeV & 
50--60\% \\
\hline
shear+bulk                 & 3000 & 1000 & 1000 & 2500 \\
shear only                 & 2000 & 1000 & 1000 & ---  \\
ideal                      & 2000 & 1000 & 500  & ---  \\
shear+bulk, no afterburner & 5500 & 5000 & 4500 & ---  \\
shear only, no afterburner & 3500 & 2000 & 1500 & ---  \\
smaller width              & 2000 & 1000 & 500  & ---  \\
no $\delta f$              & 2000 & 1000 & 500  & ---  \\
Chapman--Enskog            & 2000 & 1000 & 500  & ---  \\
\hline\hline
\end{tabular}
\label{tab:nevents_summary}
\end{table*}

\subsection{Ensemble-level analysis}

The second stage is a function that aggregates the outputs from all events to compute the final, ensemble-averaged observables. This function first computes global quantities such as the ensemble-averaged mean $p_T$ and the standard deviation of the mean $p_T$ fluctuations. It then computes the final $v_0(p_T)$ curves for each method and particle species by averaging per-event values, which allows for a robust estimation of statistical uncertainties. This implementation uses the following formulas for the two methods (see Sec.~\ref{sec:method}):

\textit{Direct covariance form:} The final result is the ensemble average of the per-event covariance, normalized by ensemble-averaged quantities. Specifically, this study computes:
\begin{equation}
    v_0^{\text{(cov)}}(p_T) = \frac{\langle n_A(p_T) \cdot [p_T]_B \rangle - \langle n_A(p_T) \rangle \langle [p_T]_B \rangle}{\langle n_A(p_T) \rangle \cdot \sigma_{[p_T]}}\,,
\end{equation}
where the standard deviation of the mean transverse momentum is calculated using the covariance between subevents, $\sigma_{[p_T]} = \sqrt{\langle [p_T]_A [p_T]_B \rangle - \langle [p_T]_A \rangle \langle [p_T]_B \rangle}$.

\textit{Symmetrized fluctuation form:} This method is implemented with a symmetrized numerator to reduce statistical biases. The final result is the ensemble average of the per-event values, defined as:
\begin{equation}
    v_0^{\text{(fluct)}}(p_T) = \frac{\left\langle \frac{1}{2}\left( \delta n_A(p_T) \delta[p_T]_B + \delta n_B(p_T) \delta[p_T]_A \right) \right\rangle}{\langle n(p_T) \rangle \cdot \sigma_{\delta [p_T]}}\,.
\end{equation}
Here, $\langle n(p_T) \rangle = \frac{1}{2}(\langle n_A(p_T) \rangle + \langle n_B(p_T) \rangle)$ is the symmetrized mean particle yield, and $\sigma_{\delta[pT]} = \sqrt{\langle \delta[p_T]_A \delta[p_T]_B \rangle}$ is the standard deviation of the mean $p_T$ fluctuations.

While these two methods are mathematically equivalent by definition, their distinct implementations can lead to minor numerical differences (see Fig.~\ref{fig:method_species}). This analysis confirms that they yield compatible results within statistical uncertainties.

\ld{
\section{Event statistics and uncertainty estimation}\label{app:events}

This appendix summarizes the number of simulated events and the method used for estimating statistical uncertainties.  
All analyses in this work are performed on an \emph{event-by-event} basis.  
For each model configuration---defined by a specific set of medium parameters, centrality class, and beam energy---this study simulates a sufficient number of hydrodynamic events to achieve reliable statistical precision.  
This enables the direct computation of event-wise observables such as particle yields and mean transverse momentum, from which the correlations entering \(v_0(p_T)\) are evaluated.  
The approximate numbers of unique hydrodynamic events are listed in Table~\ref{tab:nevents_summary}.  
To further enhance the statistics, each hydrodynamic event is oversampled 20 times and subsequently propagated through the hadronic afterburner.

The numbers in Table~\ref{tab:nevents_summary} are rounded for readability.  
The exact counts vary slightly for several practical reasons: 
for instance, the total number of requested events often exceeded the number completed within the allocated wall time, and the number of events required to achieve comparable statistical precision varies with multiplicity across different collision systems and model configurations.

The uncertainties shown in the model calculations are purely statistical.  
They are estimated using a jackknife-like procedure, in which the sample variance of the event-by-event distributions of particle yields and mean \(p_T\) (computed independently for the two subevents) is propagated to the final \(v_0(p_T)\) observable.  
This procedure naturally accounts for event-by-event fluctuations in both multiplicity and transverse momentum, providing a consistent and transparent estimate of the statistical uncertainty.

}

\bibliography{refs}
\end{document}